\documentclass{article}
\def\p{\partial}
\def\s{\sigma}

\def\de{\delta}
\def\De{\Delta}
\def\ld{\lambda}
\def\Ld{\Lambda}

\def\e{\eta}

\def\rh{\rho}

\def\b{\beta}

\def\a{\alpha}

\def\pdellx'{\frac{\partial}{\partial x'}}
\def\pdellw'{\frac{\partial}{\partial w'}}
\newcommand{\be}{\begin{equation}}
\newcommand{\ee}{\end{equation}}
\def\bed{\begin{displaymath}}
\def\eed{\end{displaymath}}
\def\bea{\begin{eqnarray}}
\def\eea{\end{eqncrray}}
\def\[{$$}
\def\]{$$}

\begin{document}

\title{The S-matrix and ghost fields \\   
in quantum Yang-Mills gravity} 
%\vspace{0.3in}
%\bigskip
\author{ Jong-Ping Hsu \\  
Department of Physics,
 University of Massachusetts Dartmouth \\
 North Dartmouth, MA 02747-2300, USA \\
E-mail: jhsu@umassd.edu}

% Beginning of the text

\maketitle
{\small  The S-matrix in quantum  
Yang-Mills gravity with translation gauge 
symmetry in flat space-time is 
investigated.  We obtain the 
generating functional of Green's functions, i.e., the 
vacuum-to-vacuum amplitude,
for Yang-Mills gravity.  The unitarity and gauge invariance of 
the S-matrix in a class of gauge 
conditions is preserved by massless ghost vector 
fields.}

%\noindent
\bigskip

\bigskip

%{\small PACS numbers: 11.15.-q,  \ \ 12.25.+e }

\bigskip
%\skip
%\pagenumbering{arabic}
%\newpage

\section{Introduction}
\noindent

The idea of a gravitational theory with space-time translational gauge symmetry is very interesting from the viewpoint of Yang-Mills theory.    It has attracted many authors \cite{1,2,3} because the translational symmetry implies the conservation of energy-momentum tensor, which is the source of the gravitational field.
But most formulations were based on curved space-time and, hence, the problems of quantization and energy-momentum conservation remain unsolved.
Recently, it has been shown that Yang-Mills gravity with translation gauge 
symmetry (or T(4) group) in 
flat 4-dimensional space-time leads to an `effective Riemannian 
metric tensor' in the limit of geometric optics of wave 
equations.\cite{3}  
Such a limiting effective metric tensor emerges in the 
Hamilton-Jacobi equation for light rays and classical particles.  In this sense,
classical general relativity based on Riemannian geometry in curved 
space-time may be interpreted as a classical manifestation of gauge 
fields with translation symmetry in flat space-time.  
Thus, it appears as if light rays and classical particles
in Yang-Mills gravity move in a `curved space-time with 
Riemannian geometry.'  However, the real underlying physical space-time of 
gauge fields and quantum particles 
is flat and, hence, has vanishing Riemann-Christoffel curvature tensor.
This property of T(4) gauge field in the limit of
geometric optics is essential for Yang-Mills gravity to be  
consistent with all known experiments.\cite{3}  Furthermore, the 
framework of flat space-time enables us to quantize Yang-Mills gravity with a 
well-defined and conserved energy-momentum tensor,
just as the usual gauge theory.  The graviton propagator is 
essential the same as that in the conventional theory.  However, the 
graviton coupling in 
Yang-Mills gravity turns out to be much more simpler than that in Einstein gravity. 
These results and properties  motivate further investigation 
of the Yang-Mills gravity.  

In sharp contrast to the 
electrodynamics with Abelian group U(1), Yang-Mills gravity
based on Abelian group T(4) of space-time translation 
symmetry needs ghost (or 
fictitious) particles to preserve the gauge invariance and unitarity of the S-matrix. 
The situation is similar to Yang-Mills theories with non-Abelian gauge groups.
Yang-Mills gravity appears to be a natural 
generalization of the conserved charge associated with 
the U(1) group to the conserved energy-momentum tensor 
of the space-time translation group. 
Furthermore, it has a big difference from the usual gauge 
theories with internal gauge groups. Namely, 
the T(4) gauge field in Yang-Mills gravity is not a (Lorentz) vector 
field with dimensionless coupling constant.  Rather, it
is a symmetric tensor field, $\phi_{\mu\nu}=\phi_{\nu\mu}$,
whose coupling 
constant $g$ has the dimension of length (in natural units, $c= \hbar = 
1$).  This is due to the fact that the generators 
$i\p/\p x^{\mu}$ of 
the space-time translation group has the dimension of 1/length and 
cannot be represented by dimensionless constant matrices.  Such a tensor field 
$\phi_{\mu\nu}$ may be termed `space-time gauge field.' 

Yang-Mills gravity with the external space-time translational symmetry 
group is a generalization of Yang-Mills theory with internal gauge 
group.  One can consider fiber 
bundle as the mathematical foundation of gauge theory with internal 
gauge group.\cite{4}  However, if the gauge group is external and 
non-compact, the corresponding fiber bundle is not as 
straightforward.  One should be careful because one cannot take 
for granted that the same mathematical and physical properties for 
internal gauge groups hold in general for external symmetry groups.\cite{5}  

\section{ The Translation Gauge-Invariant Action and Gauge-Fixing 
Lagrangian}
\noindent

Yang-Mills gravity can be formulated in both inertial and 
non-inertial frames and in the presence of fermion 
fields.\cite{1,2}   It is difficult to discuss quantum field 
theory and particle physics even in a simple non-inertial frame with a 
constant linear acceleration, where the accelerated transformation of 
space-time is 
smoothly connected to the Lorentz transformation in the limit of zero 
acceleration.\cite{6,7,8}  For simplicity, let us consider 
quantum Yang-Mills gravity in inertial frames with the Minkowski 
metric tensor $\eta^{\mu\nu}=(1,-1,-1,-1)$ and in the absence of 
fermions.  The  
 action $S_{pg}$ for pure gravity, involving space-time gauge 
 fields $\phi_{\mu\nu}(x)$ 
and a gauge-fixing Lagrangian, is assumed to be\cite{1}
\be
S_{pg}=\int (L_{\phi} + L_{\xi}) d^{4}x,
\ee
%%%%%%%%%%1
\be
L_{\phi}= \frac{1}{4g^2}\left (C_{\mu\nu\a}C^{\mu\nu\a}- 
2C_{\mu\a}^{ \ \ \  \a}C^{\mu\b}_{ \ \ \  \b} \right), 
\ee
%%%%%%%%%%%%2
$$C^{\mu\nu\a}= J^{\mu\s}\p_{\s} J^{\nu\a}-J^{\nu\s} 
\p_{\s} J^{\mu\a}, \ \ \ \ \  J_{\mu\nu}=\eta_{\mu\nu}+ 
g \phi_{\mu\nu} = J_{\nu\mu} , $$
where $C^{\mu\a\b}$ is the T(4)  gauge curvature and $c=\hbar=1$.  
We note that the Lagrangian
 $L_{\phi}$ changes only by a divergence under the translation gauge 
 transformation, and the action 
functional $S_{\phi}=\int L_{\phi} d^{4}x$ is 
invariant under the space-time translation 
gauge transformation.\cite{1}  To quantize Yang-Mills gravity, it is
necessary to include a gauge fixing 
Lagrangian $L_{\xi}$ in the action functional (1).  For example, the 
gauge fixing Lagrangian enables us to have a well-defined graviton 
propagator (see eq. (23) below).
 The gauge-fixing Lagrangian 
$L_{\xi}$ is assumed to be 
\be
L_{\xi}=\frac{\xi}{2g^{2}}\left[\p^\mu J_{\mu\a} - 
\frac{1}{2} \p_\a J \right]\left[\p_\nu J^{\nu\a} - 
\frac{1}{2} \p^\a J \right],
\ee
%%%%%%%%%%%%4%%%%%%%3%%%%%4%%%5%%%%3
\be
J=J^{\ld}_\ld=\delta^{\ld}_{\ld}- g\phi, \ \ \  \phi=\phi^{\ld}_\ld,  
\ee
%%%%%%%%%1%%%%%%%%%2%%%%%5%%%%%%%%%%4%%%%5%%%%%%%%%%%%%%%%%%%%%4
where $L_{\xi}$ involves an arbitrary gauge parameter 
$\xi$.  The Lagrangian in (3) corresponds to
a class of gauge conditions of the following form,
\be
\frac{1}{2}[\e^{\mu\rho}\e^{\nu\ld}+\e^{\nu\rho}\e^{\mu\ld} 
-\e^{\mu\nu}\e^{\rho\ld}]
\p_{\ld}J_{\mu\nu}=\p_{\ld} J^{\rho\ld} - \frac{1}{2} \p^{\rho} J 
= Y^{\rho}, 
\ee
%%%%%%%%%%6%%%%%%%%%%%%5%%%%%%%%%3%%%%%%%%%%%%%%%%%%%%5
where $Y^{\rho}$ is a suitable function of space-time.

The Lagrangian for pure gravity $L_{pg}=L_{\phi}+
L_{\xi}$ can be expressed in terms of space-time gauge fields $\phi_{\mu\nu}$:
\be
L_{pg}= L_{2}+L_{3} +L_{4} +L_{\xi},
\ee
%%%%%%%%%%%%%7%%%%%%6
where
\be
L_{2}=\frac{1}{2}\left(\p_{\ld} \phi_{\a\b} \p^{\ld} \phi^{\a\b}\right.-
\p_{\ld} \phi_{\a\b} \p^{\a} \phi^{\ld\b}-
\p_{\ld} \phi\p^{\ld} \phi \ \ \ \ \ \ \ \ 
\ \ \ \ \
\ee
%%%%%%%%%%%%8%%%%%%%%7
$$+2\p_{\ld} \phi \p^{\b} \phi^{\ld}_{\b}- 
\p_{\ld} \phi^{\ld\mu} \p^{\b} \phi_{\mu\b}),$$
 
\be
L_{\xi}=\frac{\xi}{2}\left[(\p_{\ld}\phi^{\ld\a})\p^{\rh}\phi_{\rh\a}
- (\p_{\ld}\phi^{\ld\a})\p_{\a}\phi +
\frac{1}{4}(\p^{\a} \phi)\p_{\a} \phi\right].
\ee
%%%%%%%%%%%%%%%%8
The Lagrangians $L_{2}$ and $L_{\xi}$ involve quadratic tensor 
field and determine the propagator of the 
graviton in Yang-Mills gravity.  The Lagrangians $L_{3}$ and $L_{4}$ correspond 
to the interactions of 
3- and 4-gravitons respectively.  They can also be obtained from 
the Lagrangian (2).

\section{Gauge Conditions and Effective Lagrangian of Yang-Mills 
Gravity}
\noindent

To quantize a field with gauge symmetry in a covariant formulation, 
one has to impose a gauge condition. In Yang-Mills gravity, it is 
non-trivial to impose a gauge condition in general because the gauge 
condition does not hold for all time.  We know that if one imposes a 
gauge condition in quantum electrodynamics (QED), the gauge condition 
satisfies a free field equation and, hence, hold for all times.  
However, this is true if and only if the gauge condition is linear.  
We have examined the problem of unitarity in QED if we imposed 
a quadratic gauge condition, we found that the gauge 
condition does not hold for all times.\cite{9}  Roughly speaking, the 
longitudinal and time-like photons are no longer free particles, their 
interaction in the intermediate steps of a physical process will 
create extra unwanted amplitudes to upset gauge invariance and 
unitarity of the S-matrix in QED.  Similar to the approach of Faddeev 
and Popov,\cite{10,11} QED with a quadratic gauge condition can be 
described by an effective Lagrangian which involves a 
ghost particle.  This ghost particle produces extra amplitudes to 
cancel those of unphysical (longitudinal and time-like) photons, so 
that the gauge invariance and unitarity of S-matrix in QED are 
restored.\cite{9}
Similar mechanism of cancellation occurs in any theory with gauge 
symmetry or distorted gauge symmetry,\cite{11,12} and in Yang-Mills 
gravity.

We follow Faddeev and Popov's approach to discuss how to fix a gauge for 
all times with the help of path integrals and derived 
the effective Lagrangian for quantum Yang-Mills 
gravity.\cite{10,11}  From the action (1) with the Lagrangian (2) and the 
gauge-fixing terms (3), we 
derived the Yang-Mills field equation
\be
H^{\mu\nu} + \xi A^{\mu\nu} = 0,
\ee
%%%%%%%%%%%%%26%%%%%%%19%%11%%%%9
$$
H^{\mu\nu} \equiv \left[\frac{}{}\p_{\ld} (J^{\ld}_{\rho} C^{\rho\mu\nu} - J^{\ld}_{\a} 
C^{\a\b}_{ \ \ \ \b}\eta^{\mu\nu} + C^{\mu\b}_{ \ \ \ \b} J^{\nu\ld}) \right.
$$
%%%%%%%%%%%%%%28%%%%%%%%%21%%%10
\be
\left. - C^{\mu\a\b}\p^{\nu} J_{\a\b} + C^{\mu\b}_{ \ \ \ \b} \p^{\nu} J^{\a}_{\a} -
 C^{\ld\b}_{ \ \ \ \b}\p^{\nu} J^{\mu}_{\ld}\frac{}{}\right]_{(\mu\nu)},
 \ee
\be
A^{\mu\nu} = \left[ \p^{\mu}\left(\p_{\ld} J^{\ld\nu} - \frac{1}{2} \p^{\nu}J \right)
- \frac{1}{2}\eta^{\mu\nu}\p^{\ld}\left(\p^{\s} J_{\s\ld} - \frac{1}{2} \p_{\ld}J\right)\right]_{(\mu\nu)},
\ee
%%%%%%%%%%%%%27%%%%%%%20%%%%%%%%12%%%%%%11
where $[...]_{(\mu\nu)}$ denotes that $\mu$ and $\nu$ in $[...]$ 
should be made symmetric.  The two terms in (11) are 
gauge-fixing terms, which are non-invariant under gauge 
transformations, similar to that in Einstein gravity.\cite{13}.

Let us consider a general class of the gauge conditions given in 
(5), where $Y^{\a}(x)$ is a suitable function independent of the 
fields and the gauge function $\Ld^{\a}(x)$.\cite{1}  With such a gauge 
condition, the vacuum-to-vacuum amplitude of the pure Yang-Mills 
gravity is given by
$$
W(Y^{\a}) = \int d[\phi^{\rh\s}] exp\left(i \int d^{4}x 
(L_{\phi}+\phi^{\mu\nu}j_{\mu\nu}) \right)
$$
\be
\times  \ det Q \ \prod_{\a}
\de(\p_{\ld} J'^{\ld\a} - \frac{1}{2} \p^{\a} J'- Y^{\a}), 
\ee
%%%%%%%%12
where $j_{\mu\nu}$ are external sources and $J'^{\mu\nu}$ are the 
T(4) gauge transformations of $J^{\mu\nu}$,\cite{1}
\be
J'^{\mu\nu}=J^{\mu\nu}-\Ld^{\ld}\p_{\ld} J^{\mu\nu} 
+J^{\ld\nu}\p_{\ld} \Ld^{\mu} + J^{\mu\ld}\p_{\ld} \Ld^{\nu}.
\ee
%%%%13
The delta function $\de(\p_{\ld} J'^{\ld\a} - \frac{1}{2} \p^{\a} J'- Y^{\a})$
in the path integral (12) is to maintain the gauge 
condition for all times.\cite{10,12,13}  The functional determinant $det Q$ 
is defined by\cite{11,12}
\be
\frac{1}{det Q}
= \int d[\Ld^{\rh}(x)]  \ \prod_{\a} \de\left(\p_{\ld} J'^{\ld\a}(x) - \frac{1}{2} \p^{\a} J'(x) -Y^{\a}(x)\right) 
\ee
%%%%%%%%%%%%%%%14
The matrix $Q$ is obtained by considering the T(4) gauge transformation of 
the gauge condition $Y^{\a}=\p_{\ld} J^{\ld\a} - \frac{1}{2} \p^{\a}J$:
$$
\p_{\ld} J'^{\ld\a} - \frac{1}{2} 
\p^{\a} J' = \p_{\ld} J^{\ld\a} - \frac{1}{2} \p^{\a}J
$$
\be
+ \p_{\mu}[-(\p_{\ld}J^{\mu\a}) \Ld^{\ld} + J^{\ld\a}\p_{\ld} \Ld^{\mu} 
+ J^{\mu\ld}\p_{\ld} \Ld^{\a}] + \frac{1}{2}\p^{\a} [(\p_{\ld}J) 
\Ld^{\ld}].
\ee
%%%%%%%15
The equation of the ghost vector field $\chi^{\a}$ can be obtained 
from $\p_{\ld} J'^{\ld\a} - \frac{1}{2} 
\p^{\a} J'-Y^{\a}=0$ with the replacement $\Ld^{\a} \to 
\chi^{a}$.\cite{9,14}  
Thus, we have
\be
\p_{\mu}[-(\p_{\ld}J^{\mu\a}) \chi^{\ld} + J^{\ld\a}\p_{\ld} \chi^{\mu} 
+ J^{\mu\ld}\p_{\ld} \chi^{\a}] + \frac{1}{2}\p^{\a} [(\p_{\ld}J) 
\chi^{\ld}]= 0.
\ee
%%%%16
Since $J^{\mu\ld}=\eta^{\mu\ld} + g\phi^{\mu\ld}$, equation (16) can 
be written as
$$
\p_{\ld}\p^{\ld}\chi^{\a}+\p^{\a}\p_{\mu}\chi^{\mu}
- g\p_{\mu}[(\p_{\ld}\phi^{\mu\a})\chi^{\ld}]
$$
\be
+ g\p_{\mu}(\phi^{\ld\a}\p_{\ld}\chi^{\mu})
+ g\p_{\mu}(\phi^{\mu\ld}\p_{\ld}\chi^{\a})
+ g\frac{1}{2}\p^{\a}[( \p_{\ld}\phi) \chi^{\ld}] = 0.
\ee
%%%%%%%%%%%17
This equation can be derived from the following Lagrangian for ghost 
fields $\chi'^{\a}$ and $\chi^{\a}$,
$$
L_{\chi'\chi}= \chi'_{\a}[ \p_{\ld}\p^{\ld}\chi^{\a}+\p^{\a}\p_{\mu}\chi^{\mu}
- g\p_{\mu}((\p_{\ld}\phi^{\mu\a})\chi^{\ld}) 
$$
\be
+ g\p_{\mu}(\phi^{\ld\a}\p_{\ld}\chi^{\mu})
+ g\p_{\mu}(\phi^{\mu\ld}\p_{\ld}\chi^{\a})
+ g\frac{1}{2}\p^{\a} ((\p_{\ld}\phi) \chi^{\ld}) ].
\ee
%%%%%%%%%%%%%18
The matrix Q in (14) can be obtained by expressing the ghost Lagrangian (18)
in the following form, 
\be
L_{\chi'\chi}= \chi'^{\mu}Q_{\mu\nu}\chi^{\nu},
\ee
%%%19
where we find that 
$$
Q_{\mu\nu}=\eta_{\mu\nu}\p_{\a}\p^{\a} + 
\p_{\mu}\p_{\nu} - g\eta_{\s\mu}(\p_{\rh}\p_{\nu}\phi^{\rh\s})+ 
g\eta_{\s\mu}\phi^{\rh\s}\p_{\nu}\p_{\rh}
$$
\be
+g(\p_{\s}\phi^{\s\rh})\eta_{\mu\nu}\p_{\rh} + 
g\eta_{\mu\nu}\phi^{\s\rh}\p_{\s}\p_{\rh} + 
g\frac{1}{2}(\p_{\mu}\p_{\nu}\phi) + 
g\frac{1}{2}(\p_{\nu}\phi)\p_{\mu}.
\ee
%%%%%%%%%%%%20
Since $W(Y^{\a})$ is invariant under an infinitesimal change of 
$Y^{\a}(x)$ for all $Y^{\a}(x)$,\cite{12} we may write 
$W(Y^{\a})$ in (12) as
$$
W= \int W(Y^{\a}) exp\left[-i\int d^{4}x \frac{\xi}{2g^{2}}Y^{\a}(x) 
Y_{\a}(x) \right] d[Y^{\a}(x)]
$$
%%%%%
$$ 
= \int d[\phi^{\s\rh}](detQ) exp \left\{i\int d^{4}x 
\left[L_{\phi}+\frac{}{}\phi^{\mu\nu}j_{\mu\nu} \right.\right.
$$
\be
- \frac{\xi}{2g^{2}}\left(\p_{\ld} J^{\ld\a} - \frac{1}{2} \p^{\a}J \right) 
 \left.\left.\left(\p_{\s}J^{\s\b}-\frac{1}{2}\p^{\b}J\right)\eta_{\a\b}\right]\right\}
\ee
%%%%%%%%%%%%%21
to within unimportant multiplicative factors.  The amplitude $W$ is 
equivalent to the following effective Lagrangian\cite{9,12}
\be
L_{eff}=L_{\phi} + \frac{\xi}{2g^{2}}\left(\p_{\ld} J^{\ld\a} - 
\frac{1}{2} \p^{\a}J\right)
\left(\p_{\s} J^{\s\b} - \frac{1}{2} \p^{\b}J\right)\eta_{\a\b} + 
L_{\chi'\chi},
\ee
%%%%%22
where the ghost Lagrangian $L_{\chi'\chi}$ is given by (19) and (20).

This effective Lagrangian $L_{eff}$ completely specifies the quantum 
Yang-Mills gravity, including the physical tensor gauge field, 
together with the unphysical vector-field $\p_{\mu}\phi^{\mu\nu}$.
and the ghost vector-fields $\chi'^{\mu}(x)$ and $\chi^{\mu}(x)$.  
Thus, the Feynman rules of quantum Yang-Mills gravity can be derived from 
the effective Lagrangian (22).

\section{Propagators of gravitons and ghost particles and their couplings}
\noindent

In general, the propagators of graviton and ghost particles and their 
couplings in Feynman rules depend on the 
specific form of gauge condition and the gauge 
parameters.  We have chosen the gauge condition
$\p_\mu J^{\mu\nu} - 
(1/2) \p^\nu J = Y^{\nu}$ with an arbitrary $\xi$ in the 
effective Lagrangian (22).  From the free Lagrangians (7) and (8) for the tensor 
gauge field $\phi^{\mu\nu}$, one can obtain the graviton
propagator,  
\be
G_{\a\b,\rho\s}= i\left[\frac{1}{k^{2}} (\e_{\a\rho}\e_{\b\s}\right.+ 
\e_{\b\rho}\e_{\a\s} - \e_{\a\b}\e_{\rho\s})
\ee
%%%%%%%%%%%%%%%%%33%%523
$$-\frac{1}{k^{4}}\left(1-\frac{2}{\xi}\right)
(\e_{\rho\a}k_{\s}k_{\b} + \e_{\rho\b}k_{\s}k_{\a} + \e_{\s\a}k_{\rho}k_{\b} 
+\left. \e_{\s\b}k_{\rho}k_{\a}){\frac{}{}}\right], $$
for arbitrary $\xi$.  This is consistent with that 
obtained by Fradkin and Tyutin because their linearized field equations 
are the same.\cite{13,15,16,17}

On the other hand, if one chooses another gauge condition
$\p_{\ld} J^{\rho\ld} = Y_{1}^{\rho}$ with an arbitrary $\xi$, i.e., the 
gauge-fixing Lagrangian (8) is replaced by
$L'_{\xi}=(\xi/2)(\p_{\ld}\phi^{\ld\a})\p^{\rh}\phi_{\rh\a}$. 
We obtain a different graviton propagator $G'_{\a\b,\rho\s}$: 
\be  
G'_{\a\b,\rho\s}= i\left[\frac{1}{k^{2}} (\e_{\a\rho}\e_{\b\s}\right.+ 
\e_{\b\rho}\e_{\a\s} - 
\e_{\a\b}\e_{\rho\s}) 
\ee
%%%%%%%%%%%%%%%%%%%%%%%%34%%%24
$$
-\frac{1}{k^{4}}\left(1-\frac{2}{\xi}\right)
(\e_{\rho\a}k_{\s}k_{\b} + \e_{\rho\b}k_{\s}k_{\a} + \e_{\s\a}k_{\rho}k_{\b} 
+ \e_{\s\b}k_{\rho}k_{\a}) 
$$
$$
+ \frac{1}{k^{4}}(\e_{\rho\s}k_{\a}k_{\b}+ \e_{\a\b}k_{\rho}k_{\s})
+ \left.\frac{1}{k^{6}}\left(1 -\frac{6}{\xi}\right)k_{\rho}k_{\s}k_{\a}k_{\b}\right].
$$  
Similarly, the propagator of the ghost particle can be obtained from 
its free Lagrangian, i.e., (19) with (20) and $g=0$.  We obtain
\be
G^{\mu\nu}=\frac{-i}{k^{2}}
\left(\e^{\mu\nu}-\frac{k^{\mu} k^{\nu}}{2k^{2}}\right).
\ee
%%%%%%%%%%%%%31%%%525
The $i\epsilon$ prescription for the Feynman propagators (23), (24) and 
(25) is understood.

In order to obtain the Feynman rule for ghost-ghost-graviton vertex, 
it is more convenient to use an equivalent form of the ghost 
Lagrangian:
$$
L_{\chi'\chi}= -\p_{\ld}\chi'_{\a}\p^{\ld}\chi^{\a}- \p^{\a}\chi'_{\a}\p_{\mu}\chi^{\mu}
+ g(\p_{\mu}\chi'_{\a})(\p_{\ld}\phi^{\mu\a})\chi^{\ld}  
$$
\be
- g(\p_{\mu}\chi'_{\a})\phi^{\ld\a}\p_{\ld}\chi^{\mu}
- g(\p_{\mu}\chi'_{\a})\phi^{\mu\ld}\p_{\ld}\chi^{\a}
- g\frac{1}{2}(\p^{\a}\chi'_{\a}) (\p_{\ld}\phi) \chi^{\ld}.
\ee
%%%%%%%%%%%%%%%%26
This equivalent form can be obtained by considering the action for 
the Lagrangian $L_{\chi'\chi}$ in (18) and using integration by parts.
The Lagrangian (26) implies that the ghost-ghost-graviton 
vertex (denoted by $\chi'^{\mu}(p)\chi^{\nu}
(q)\phi^{\a\b}(k)$) is
\be
 \ ig\left[-p^{\a}k^{\mu}\eta^{\nu\b} +
p^{\mu}q^{\a}\eta^{\nu\b} + p^{\a}q^{\b}\eta^{\mu\nu}
+\frac{1}{2}k^{\mu} p^{\nu} \eta^{\a\b}\right]_{(\a\b)},
\ee
%%%%%31%%%%%%%%%%%%%%%%%%%%%%%%%%%%%%%%%32%%26%%%%%%%%%%%%27
where $[...]_{(\a\b)}$ denotes that the indices $\a$ and $\b$ in $[...]$ should be made 
symmetric.  In the expression (27), all momenta are
incoming to the vertex, $p_{\ld}+q_{\ld}+k_{\ld}=0$. 
  All ghost-particle vertices are 
bilinear in the ghost particle, as shown in the ghost Lagrangian 
(19).  As a result in the Feynman rules, the ghost particles appears, 
by definition of the physical subspace for the S-matrix, only in closed loops in the 
intermediate states of a physical process.  Moreover, 
there is a factor of -1 for each ghost particle loop.\cite{9,11}
Thus, the ghost particle resembles a fermion in the Feynman rules.  
The translation gauge invariance of Yang-Mills gravity
implies that the observable results such as those 
obtained from the S-matrix
 should be independent of the gauge parameter $\xi$ in the 
 gauge-fixing term $L_{\xi}$ given in (3).\cite{11,12}

\section{Discussions}
\noindent
 
The theoretical structures of Yang-Mills gravity with external 
space-time translation group have some similarities and some 
differences 
from the usual gauge field theories.  For one thing, the generators 
of the space-time translation group T(4) do not have the constant matrix 
representation.  Consequently, the fiber bundle corresponding to 
Yang-Mills gravity is not straightforward.  The T(4) gauge covariant derivative, 
$\De^{\nu}=J^{\nu\ld}\p_{\ld}$, satisfies the Jacobi identity,
$[\De^{\ld},[\De^{\mu}, \De^{\nu}]]+[\De^{\mu},[\De^{\nu}, \De^{\ld}]]+
[\De^{\nu},[\De^{\ld}, \De^{\mu}]] \equiv 0.$  But the corresponding 
Bianchi identity for the T(4) gauge curvature $C^{\mu\nu\a}$ differs 
from that in the usual gauge theory.  We obtain a modified Bianchi identity:
$$
(\de^{\ld}_{\a}\De^{\rh}-\p^{\ld}J^{\rh}_{ \ \a})C^{\mu\nu}_{ \ \ \ \ld}
+(\de^{\ld}_{\a}\De^{\mu}-\p^{\ld}J^{\mu}_{ \ \a})C^{\nu\rh}_{ \ \ \ 
\ld}$$
\be
+ (\de^{\ld}_{\a}\De^{\nu}-\p^{\ld}J^{\nu}_{ \ \a})C^{\rh\mu}_{ \ \ \ \ld}
\equiv 0.
\ee
%%%%%27%%%%%%%%%%%%%%28
 
In usual gauge theories, a given gauge condition does not uniquely 
determine the effective Lagrangian for a theory. For example, there 
may be different ghost Lagrangians which can restore the gauge 
invariance and unitarity of the S-matrix in a 
theory.\cite{14} In some cases, when one uses the Lagrange 
multiplier, one obtains a different ghost Lagrangian, which can also 
cancel the unwanted amplitudes to restore unitarity.  These properties
appear to be true in a theory with a gauge symmetry.\cite{14,18}

In contrast to the previous works\cite{1,2},
Ning Wu recently attempted to give Einstein's gravity a 
new interpretation based on a T(4) gauge symmetry.\cite{19}  Namely, 
the conventional Lagrangian in the Hilbert-Einstein 
action\cite{20} is expressed in terms of the T(4) gauge curvature 
$C^{\mu\nu\a}$ within the framework of flat 
space-time.  Wu gave a formal proof that the quantum gravity based on 
the Hilbert-Einstein action with the T(4) gauge symmetry in flat 
space-time is renormalizable.\cite{21,22}  In view of the profound 
difficulties in the quantization of Einstein's gravity in curved 
space-time,\cite{23} it is desirable to have explicit 
calculations to substantiate a formal proof of renormalizability. 

From the viewpoint of Feynman rules, there is a significant difference 
in the structure of interaction vertices between Yang-Mills gravity 
in flat space-time 
and other theories of gravity,\cite{1,2,11,13,24,25} including Wu's 
formulation of gravity\cite{21}.  Namely, the maximum number of 
gravitons in a vertex is 4, as one can see in the total Lagrangian 
(22) of Yang-Mills gravity.  On the other hand, there exist vertices with  
arbitrary large numbers of gravitons in Einstein gravity 
and in other theories of gravity.  For an ordinary tensor field 
theory with a dimensional coupling constant and without a gauge 
symmetry, one would expect that the theory is not renormalizable 
based on power counting.  Nevertheless, this argument may not be 
applicable to Yang-Mills gravity with T(4) gauge symmetry.
This maximum 4-vertex for graviton coupling, together with the T(4) gauge symmetry
and the conserved energy-momentum tensor, may be a gateway to the renormalizable quantum 
Yang-Mills gravity.  
  
\bigskip

\noindent

{\bf Acknowledgements}
\bigskip

The author would like to thank Dana Fine for discussions of fiber 
bundle with external space-time translation group.
  The work was supported in part by  Jing Shin
Research Fund and Prof. Leung Memorial Fund of the UMass Dartmouth Foundation.

\newpage
{\bf Appendix.  \  Ghost Particles and Unitarity of the S-matrix}
\bigskip

Let us give some arguments and a proof for the unitarity of 
the S-matrix for pure gravity based on the effective Lagrangian (22).  
For a discussions of unitarity of the Yang-Mills gravity with the 
gauge condition in (5), one can write 
the ghost field $\chi^{\mu}$ in the following form,\cite{13} 
\bed
\hspace{1in} \chi^{\mu}(x) = \int d^{4}y {\bf D}^{\mu}_{\nu}(x,y,\phi_{\a\b})\hat{\chi}^{\nu}(y)
 \ \ \ \ \ \ \ \ \ \ \ \ \hspace{0.7in}  (A1)
\eed
%%%%32%%%%%%%%%%%%%19A1
where $\chi^{\mu}$ satisfies equation (17) 
which can be written in the form
\bed
\hspace{1in} Q_{\mu\nu} \ \chi^{\nu} = 0, \ \ \ \ \ \ \ \ \ \  
\hspace{2in} (A2)
\eed
%%33%%%%%%%%%%%%%%%%%%%%20A2
where $Q_{\mu\nu}$ is given in (20).   Clearly, in the limit $g \to 
0$, one has $J_{\mu\nu} \to \eta_{\mu\nu}$.  Thus, the operator
$Q_{\mu\nu}$ reduces to a non-singular differential operator in this 
limit,
\bed
\hspace{1in} Q_{\mu\nu} \to  \eta_{\mu\nu}\p_{\ld}\p^{\ld}+\p_{\mu}\p_{\nu} \equiv 
Q^{0}_{\mu\ld}. \ \ \ \ \ \ \ \   \hspace{1.1in} (A3)
\eed
%%%%%%%%%34%%%%%%%%%%%%%21A3
This limiting property can be seen 
from (20).  One can choose the function
${\bf D}^{\mu}_{\nu}(x,y,\phi_{\a\b})$ in equation (A1) to have the specific 
form
\bed
\hspace{1in} {\bf D}^{\mu}_{\nu} = \left[Q^{-1}\right]^{\mu\ld} \overleftarrow{Q}^{0}_{\ld\nu}
 \ \ \ \ \ \ \ \ \ \ \ \  \hspace{1.5in} (A4)
\eed
%%%%%%%%35%%%%%%%%%%%%%%%22A4
so that $\hat{\chi}^{\mu}$ satisfies the free field equation,
\bed
\hspace{1in} Q^{0}_{\ld\mu}\hat{\chi}^{\mu}=[ \p^{\s}\p_{\s} \eta_{\ld\mu} 
+ \p_{\ld}\p_{\mu}]\hat{\chi}^{\mu}= 0. \ \ \ \ \ \ \   \hspace{0.8in} (A5)
\eed
%%%%%36%%%%%%%%%%%%%%%%23A5

The generating functional for connected Green's functions in gauge invariant gravity can be 
defined after the gauge condition is specified.\cite{13} 
Similarly, in Yang-Mills gravity the generating functional for 
connected Green's functions (or the 
vacuum-to-vacuum amplitude) (21) can be written as\cite{12,26}
\bed
W^{Y}_{\xi}= \int d[\phi_{\a\b}]
exp\left[i\int d^{4}x\left(L_{\phi}+ 
\frac{\xi}{2g^{2}}Y^{\mu}Y_{\mu} + 
\phi^{\mu\nu}j_{\mu\nu}\right)\right.
\eed
$$
\hspace{1in} \left. + Tr \ ln \ Q(\overleftarrow{Q}^{0})^{-1}\right], \ \ \ \ \ \ 
\ \ \ \    \hspace{1.5in} (A6)
$$
%%%%%%%A6
where the external sources $j_{\mu\nu}$ are arbitrary functions and
$Y^{\mu}$ is given by equation (5).  It follows from (A5) and (A6) 
that the S-matrix corresponding to the generating functional (A6) is 
unitary.\cite{13,11}  The T(4) gauge field equations, $H^{\mu\nu}=0$, 
for pure gravity hold in the physical subspace.

The last term in (A6) can  be 
written in terms of vector-fermion ghost fields $\chi^{\a}(x)$ and 
${\chi'}^{\b}(x)$,\cite{13,26}
\bed
\hspace{0.7in}Tr \ ln  \ Q(\overleftarrow{Q}^{0})^{-1} = \int 
d[\chi^{\a},\chi'^{\b}]exp\left(i\int L_{\chi'\chi}d^{4}x\right),
   \hspace{0.23in} (A7)
\eed
%%%%%39%%%%%%%%%%%%%%26
where the effective action $L_{\chi'\chi}$ is given by (19), which  
describes the ghost particles associated with the gauge specified in 
(5). 
Note that $\chi'^{\mu}$ is considered as an independent field.
The quanta of the fields $\chi^{\mu}$ and $\chi'^{\mu}$ in the 
effective Lagrangian $L_{\chi'\chi}$ are the ghost (or fictitious)
particles in Yang-Mills gravity.  By definition of the physical 
states for the S-matrix, these ghost particles cannot exist in the 
external states.  They can only appear in the intermediate 
steps of a physical process.\cite{12,14}

%% Bibliography
%%%%%%%%%%%%%%%%%%%
%\newpage
%\section*{References}

\bibliographystyle{unsrt}

\end{document}